\documentclass[preprint,authoryear,12pt]{elsarticle}

\usepackage{amsmath}
\usepackage{epsfig}
\usepackage{natbib}
\input{journaldef.sty}

\usepackage{amssymb}

\usepackage[ps2pdf,%
a4paper=true,%
breaklinks=true,%
colorlinks=true,%
pdfauthor={First Author et al.},%
pdftitle={Template for manuscripts in Advances in Space Research}%
]{hyperref}

\journal{Advances in Space Research}

\begin{document}

\begin{frontmatter}



\title{Evolution of electron beam pulses of short duration in the solar corona}


\author{G. A. Casillas-P\'erez\corref{gil}}
\address{Instituto de Geof\'isica, Universidad Nacional Aut\'onoma de M\'exico, 
Ciudad Universitaria, C.P. 04510, Ciudad de M\'exico, M\'exico.}
\cortext[gil]{Corresponding author}
\ead{gacp@geofisica.unam.mx}


\author{S. Jeyakumar}
\address{Departamento de Astronom\'ia,  Universidad de Guanajuato, M\'exico,
Guanajuato, Gto., C.P. 36000, M\'exico.}

\author{H. R. P\'erez-Enr\'iquez}
\address{Centro de Geociencias, Universidad Nacional Aut\'onoma de M\'exico, 
Campus Juriquilla, C.P. 76230 Juriquilla, Quer\'etaro, M\'exico.}

\author{M. A. Trinidad}
\address{Departamento de Astronom\'ia,  Universidad de Guanajuato, M\'exico,
Guanajuato, Gto., C.P. 36000, M\'exico..}

\begin{abstract}

Narrowband radio bursts with durations of the order of milliseconds,
called spikes, are known to be associated with solar flares.
In order to understand the particle beams responsible for the radio 
spike phenomena, evolution of electron beam pulses injected from a 
solar flare region into the corona is studied. 
Numerical integration of the Fokker-Planck (FP) equation 
is used to follow the evolution of the electron beam pulse. 
The simulations show that the short duration pulses lose most of their energy 
within a second of propagation into the corona. Electron beam with
a small low energy cut off is thermalised faster than that with
a high low energy cut off. 
\end{abstract}


\end{frontmatter}

\parindent=0.5 cm

\section{Introduction}\label{s:intro}

Explosive events occurring in the Sun produce radiation in a wide
range of the electromagnetic spectrum. In the radio band, 
particularly of interest are the fast transient events known 
as radio spikes. These are events with enhanced
radio emission with durations of the order of tens of milliseconds
[\citet{Vlasov2002}; \citet{Krucker1997}, \citet{Bouratzis2015}], 
that surpass the solar radio emission of the quiet sun \citep{Benz2009}.
The study of solar radio spikes has been of interest for some decades
and has been considered to be important for understanding the 
physical processes occurring in the solar corona and their possible association 
to other large energy releasing events such as  
solar flares and coronal mass ejections [\citet{Chen2015}, 
\citet{Aschwanden1992}, \citet{Dabrowski2003},
\citet{Baolin2012}, \citet{Baolin2013}, \citet{Dabrowski2011}].
Recently, unusual radio bursts of millisecond durations were also observed to
be associated with the solar flares \citep{Oberoi2009}.

There are numerous observations of solar radio spikes in different
frequency bands [\citet{Gudel1990}; \citet{Fleishman2003}; \citet{Vlasov2002}; 
\citet{Magdalenic2006}; \citet{Rozhansky2008}; \citet{Dabrowski2011}; 
\citet{Li1986}; \citet{Shevchuk2016}]. Observations of these spikes show positive, negative, or null drift rates  
over a very narrow bandwidth, which implies 
a small source extent for these events. 
These spikes have also been observed in groups from tens to thousands 
inside broad-band radio bursts of type I, III and IV and together with X-ray bursts.  
These observations indicate that the particles responsible for the radio spikes
are also produced during reconnection events. 

Based on the observed high brightness temperature of the radio emission,
which reach values up to $10^{13} - 10^{15} K$ or higher
[\citet{Benz1986}; \citet{Dabrowski2011}; \citet{Vlasov2002};
\citet{Aschwanden1990}], the spikes necessarily have to be associated 
with a non-thermal coherent emission mechanism [\citet{Aschwanden1992};
\citet{Benz1986}; \citet{Dabrowski2011}; \citet{Vlasov2002}]. 
Both electron cyclotron maser (ECM) and plasma waves 
were proposed as candidates for the possible radio emission process.
In the case of plasma waves, the magnetic field strength is not a relevant 
factor influencing the radio emission [\citet{Melrose1975}; \citet{Zheleznyakov1975}].
Many observed features of radio spikes could be explained in the framework of the plasma waves
[\citet{Warwick1969}; \citet{Melnik2014}]. It is assumed that the radio emission from solar bursts
is initiated by instabilities in the plasma. These instabilities cause the 
generation of Langmuir waves (longitudinal) which drive the generation of radio emissions by 
means of wave-wave interactions [\citet{Tang2009}; \citet{Li2013}; \citet{Baolin2013}; 
\citet{Baolin2013}; \citet{Dabrowski2015}]. On the other hand, the ECM mechanism suggests 
that radio emission is produced by wave-particle interactions arising from loss-cone 
instabilities \citep{Aschwanden2005}. For the occurrence of the ECM emission a strong 
magnetic field is required, which can be expected only in dense inhomogeneities 
[\citet{Fleishman2003}; \citet{Kuijpers1981}]. 

Irrespective of the emission process, fundamental to the generation of spikes is
a propagating electron beam. The electrons are accelerated either during the magnetic 
reconnection or by propagating shocks in the solar flare region [\citet{Miller1997};
\citet{Zharkova2011}]. The evolution of the electrons injected into a solar magnetic 
loop  has been numerically studied in order to understand the phenomena of radio spikes 
[e.g., \citet{Aschwanden1990}].

However, the electrons that propagate away from the acceleration site towards the outer corona
produce intense radio emission with rapid drifts in frequency, known as type III radio bursts. 
Numerical simulations of these outward propagating electron beams were carried out in order to study the 
type III bursts [\citet{Ratcliffe2014}; \citet{Reid2013}; \citet{Li2008}; \citet{Li2011c}]. These 
simulations show that the onset of Langumuir wave generation and associated 
onset of radio emission depends on the duration of the electron beam injection, 
where bursts observed at frequencies greater than about 1~GHz require beam durations of 
milliseconds or lower. These simulations launch the electron beam over a wide region in space
estimated from type III observations. However in order to produce short duration spikes
it is necessary to study  the evolution of electron beam pulses of short duration 
injected in a smaller region, such as density inhomogeneities or micro structure 
on the reconnection region.  Therefore in this work the evolution of short duration electron 
beam pulse in the solar corona, injected in a small region of space, is studied numerically.
We have developed a numerical code to solve the time dependent FP equation based on the solution 
strategy of \citet{Hamilton1990}. We use the code to study the evolution of short duration 
electron beam pulses. In section~\ref{s:model} we describe the model for the evolution of the 
particle distribution function. In section~\ref{s:pulse} we describe the parameters of the 
electron beam pulse. In section~\ref{s:results}  the results obtained from simulations are 
presented. The conclusions are presented in the section~\ref{s:conclusion}.

\section{Model of electron beam evolution}\label{s:model}

The evolution of phase space distribution function $f(E, \mu, s, t)$ 
per unit volume (cm$^{-3}$), unit energy (KeV) and unit pitch angle cosine
($\mu$) of electron beams propagating in the solar corona is 
followed at different times and position. The dominant
processes that affect the energy and pitch angle of the particles are the
Coulomb collisions and magnetic mirroring. In addition, such a propagating 
electron beam also induces an electric field and a neutralizing return current
\citep{McClements1992b, Siversky2009}.

The evolution of the energetic particle phase space distribution function 
can be described  by the time dependent FP equation with the following terms \citep[e.g.,][]{lifshitz.etal1981, Hamilton1990}.

  \begin{equation}
    \label{eqn:fp}
  {
    \begin{aligned}
   \frac{\partial f}{\partial t} &
          +  \mu c \beta \frac{\partial f}{\partial s} 
          +               \frac{\partial (\dot{ \mu} f) }{\partial \mu} 
          +               \frac{\partial (\dot{ E} f) }{\partial E} 
          -               \frac{e\mathcal{E}}{m_e}\mu \frac{\partial f}{\partial v} 
          -               \frac{e\mathcal{E}}{m_e v}( 1 - \mu^2) \frac{\partial f}{\partial \mu}  \\
      &  =
       \frac{ \partial }{\partial \mu}  \left( D_{\mu\mu} \frac{ \partial f}{\partial \mu} \right) 
        + \left( \frac{\partial f}{\partial t} \right)_{mag-mirror} +  S(E, \mu, s, t)
    \end{aligned}
   }
  \end{equation}

In the above equation  $E$ is energy, $\mu$ is the cosine of the pitch angle,
$s$ is the position from the top of the coronal loop and $\beta=v/c$.
The function $S(s, E, \mu, t)$ describes the source function. 
The variable $\mathcal{E}$ is the electric field induced by the electron beam, $e$ and $m_e$ are the electron charge and mass respectively, $c$ is the speed of light,
$D {\mu\mu}$ is the diffusion coefficient due to Coulomb collisions and $v$ is
the velocity.
    
Substituting the terms $\dot{ \mu }$, $\dot{ E }$ and $D {\mu\mu}$
corresponding to the magnetic mirroring and Coulomb collisions as given in
\citet{Hamilton1990}, the above equation can be written in scaled units as,

  \begin{equation}
    \label{eqn:fp1}
  {
    \begin{aligned}
   \frac{\partial f}{\partial \tau } =  &
          -  \mu \beta \frac{\partial f}{\partial \xi } 
          +              \frac{ \beta}{2} \frac{d \ln{B} }{d\xi} \frac{\partial }{\partial \mu} \left[ \left(1 - \mu^2 \right) f \right]
          +              \eta \ln{\Lambda} \frac{\partial }{\partial E } \left( \frac{f}{\beta} \right)
\\
&
          +              \frac{ \eta \ln{\Lambda} }{\beta^2 \gamma^2 }  \frac{\partial }{\partial \mu }  \left[ \left( 1 - \mu^2 \right) \frac{\partial f}{\partial \mu } \right]
          +               \alpha \varepsilon \mu \beta  \frac{\partial f}{\partial E} 
          +                \alpha \frac{ \varepsilon ( 1 - \mu^2)}{\beta} \frac{\partial f}{\partial \mu}  \\
      &   +  S(E, \mu, s, t)
    \end{aligned}
   }
  \end{equation}

Here $E$ is expressed in units of electron rest mass energy, and other scaled variables are given below.

\begin{eqnarray*}
 \tau_c =  1/(4\pi n_0 c r_0^2) 
~~~
 \tau = t/\tau_c   
~~~
 \xi = s / (c \tau_c)   
~~~
 \eta = n/n_0  
~~~
 \varepsilon = \mathcal{E}/\mathcal{E}_0  
\\
 \alpha = \frac{ e \mathcal{E}_0 \tau_c}{m_e c}  
~~~
 \mathcal{E}_0 =  \frac{m_e c^2}{2\pi e^3 n_0}
~~~
\end{eqnarray*}

\noindent where $\ln{\Lambda}$ is the Coulomb logarithm which is assumed to be $\approx
10$, $r_0$ is the classical radius of the electron, ${n_0}$ is unit of ambient density scale and 
the variable $n$ is the ambient density. The numerical schemes implemented for the
energy, position and $\mu$ terms of the FP equation are based on the strategy proposed 
by \citet{Hamilton1990}. The electric field $\mathcal{E}$ can be calculated using equation~3 of 
\citet{ZharkovaGordovskyy2006}. 
However, the results of the simulations reported in this work do not include the 
effects of the magnetic convergence nor the electric field terms, which corresponds to those with
the variables $\varepsilon$ and $B$ on the right side of equation \ref{eqn:fp1}. 

\section{Simulation of electron beam pulse propagation}\label{s:pulse}

The evolution of an electron beam pulse is studied by launching 
a pulse of short duration from the top of a magnetic loop into the solar corona.
Based on the observations of Xray emission from precipitating electron beams,
the duration of the pulse is of the order of $10^{-3}$~s 
{[\citet{Charikov2004}; \citet{Siversky2009}]. 
The pulse is injected at a height of $10^9$~cm from the photosphere and 
the evolution is followed to a distance of about $3 \times 10^{11}$~cm.

\begin{table}
\centering
\caption{Parameters of the  simulation runs 
\label{tab:simpars}}
\begin{tabular}{l c c c}
\hline
 Run &  $\gamma$  &  $E_0$~(Kev)  &  $n_b$~cm$^{-3}$  \\
\hline
 B1  &   3        &   16        &  $1.75 \times 10^7$  \\
 B2  &   3        &    7        &  $1.74 \times 10^7$  \\
 B3  &   7        &   16         &  $2.87 \times 10^7$  \\
\hline
\end{tabular}
\end{table}

The electron beam is injected at the lower boundary using the boundary 
condition for the electron distribution function of the form,

\begin{figure}[t]
\hbox{
\includegraphics[width=6.5cm, bbllx=86pt, bblly=55pt, bburx=364pt, bbury=300pt]{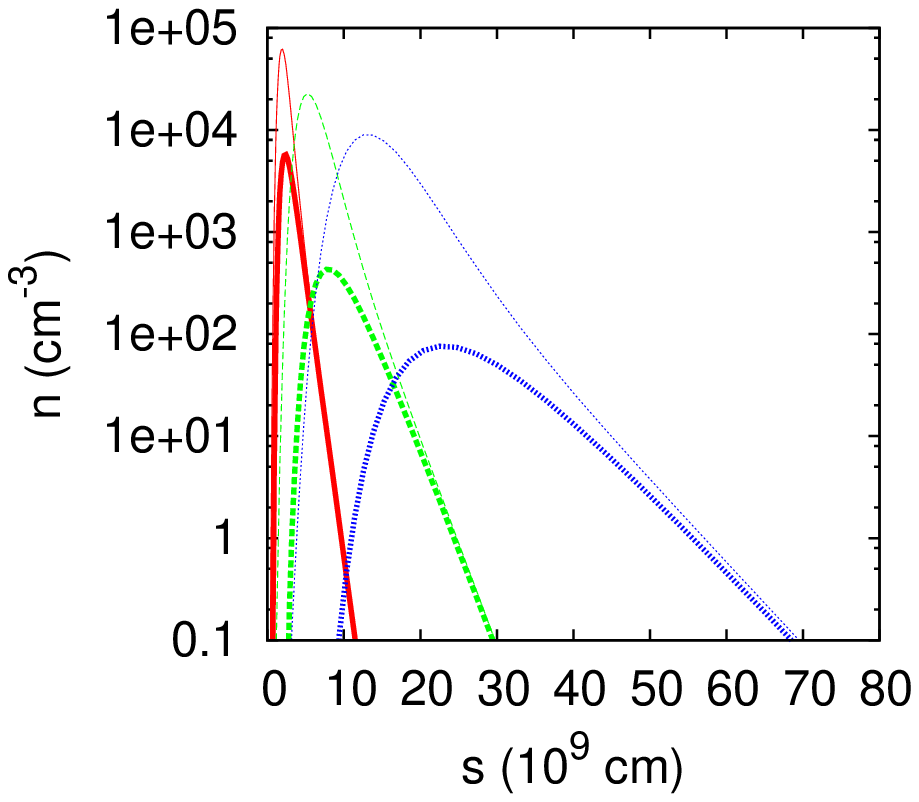}
\includegraphics[width=6.5cm, bbllx=86pt, bblly=55pt, bburx=364pt, bbury=300pt]{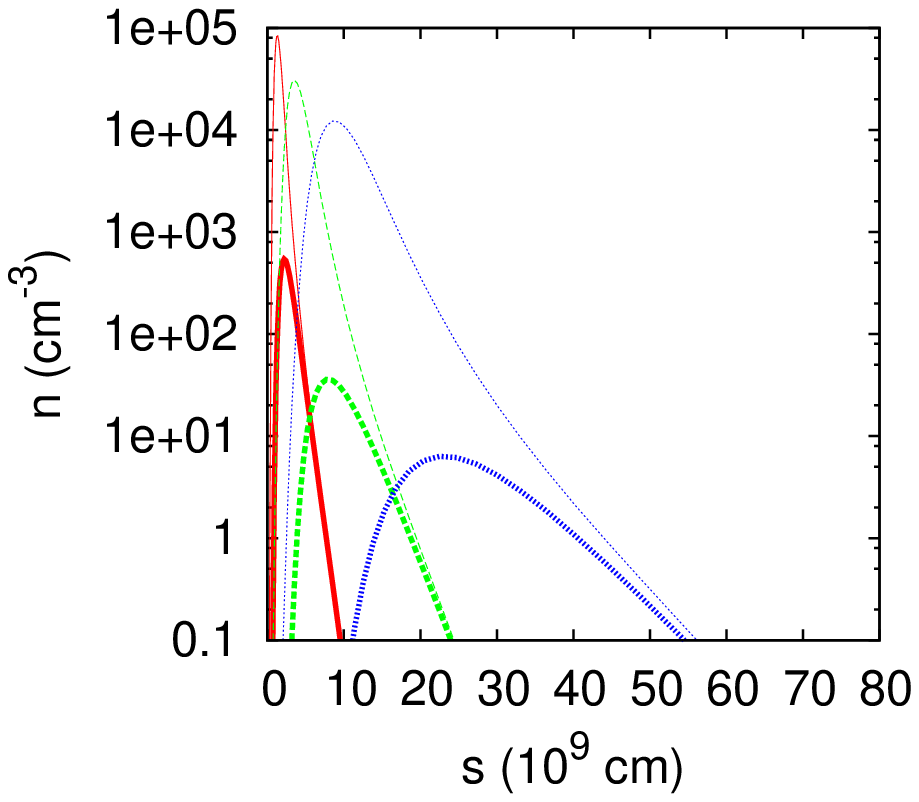}
}
\centering

\caption{The total density n~$(cm^{-3})$ as a function of position is plotted
at times $t=0.282,0.785$ and $1.968~s$ as solid (red), dashed (green) and dotted (blue)
curves respectively. The thin lines show results from model runs with only positional
advection and the thick lines show that of the model runs including Coulomb collisions.
The left and right panels correspond to model run B1 ($E_0$ = 16 Kev) and B2 ($E_0$ = 7 Kev)
respectively.
}\label{fig:tdenpos}
\end{figure}

\begin{figure}
\includegraphics[width=6.5cm, bbllx=86pt, bblly=55pt, bburx=364pt, bbury=300pt]{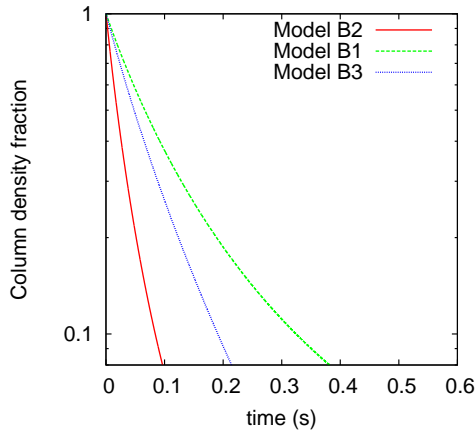}
\centering
\caption{The ratio of column density through the simulation domain
of model runs B with those of A are plotted against time.
The dashed (green), solid (red) and dotted (blue) curves represent model runs 
B1, B2, and B3 respectively. 
The loss of the electron population can be seen as decreasing fraction over time.
}\label{fig:tdenvariation}

\end{figure}

\begin{equation}\label{eqn:fdist1}
f(E,\mu)=f_n\psi(\tau) \frac{(E/E_0)^{\delta - 1} }{(E/E_0)^{\delta + \gamma} + 1} \textnormal{exp}(-\frac{(1-\mu)^2}{\Delta \mu^2})
\end{equation}
where $\delta=10$, is the power-law index of the electron distribution 
below the cut off energy of $E_0$, and $\gamma=3$ is the value of the power-law 
index for energies above $E_0$, $\Delta\mu=0.2$ is the Gaussian width of the
pitch angle distribution. The range of energy of the distribution is 
$E_{min}= 1.2$~KeV and $E_{max}=384$~KeV, with a cutoff energy of 
$E_0=16$ or $7$~KeV.  The temporal profile $\psi(\tau)$ of the beam, is 
taken as rectangular function with a width of $0.2 \times 10^{-3}$~s. 
The beams are injected at the lower boundary, which corresponds to a 
physical size of about 100~km along the propagation direction, for the 
resolution used in the simulations.

The initial condition of the electron beam distribution is 
$f(\tau=0, \xi,E,\mu)=0$, 
and the upper boundary condition is
$f(\xi_{max},E,\mu)=f(\xi_{max}-d\xi,E,\mu)$.
The ambient density profile in the corona is described  \citep[e.g.][]{Ratcliffe2013},

\begin{equation}\label{eqn:densratcliffe}
n=n_0 ~ \textnormal{exp}\left( - (s - s_0)/H_{scale} \right)
\end{equation}

\noindent where $n_0 = 10^{10}$~cm$^{-3}$ is the density at the lower boundary and 
$H_{scale} = 10^{10}$~cm is the scale height. 
For the range of distances where the electron beam is followed, 
the variation of the magnetic field is not that significant. 
Moreover, the effects of induced electric field on the evolution of
the electron beam pulses of short duration is small \citep{Siversky2009}. 
Therefore as a first approximation, the simulations presented here
do not include the magnetic convergence and the electric field terms.
The simulations were carried out with a resolution of 100, 30 and 30 grids
in position,  $\mu$ and energy  axes respectively. The position and energy axes
were divided logarithmically. The simulations runs were carried out for 
different low energy cut offs (B1 and B2)  and high energy spectral index (B3). 
For comparison, simulations excluding all but positional affection terms 
were also carried out (runs A1, A2, and A3 respectively). The distribution function is normalized 
with a flux at the lower boundary, $F_0 = 10^{12}$~erg~cm$^{-2}$~s$^{-1}$. 
The parameters used for the simulations and the effective total density are given
in table~\ref{tab:simpars}

\section{Results and Discussion}\label{s:results}

\subsection{Effect of collisions on the propagating pulse}\label{s:collision}

The variation of the density of the electrons with position is shown in
the Figure~\ref{fig:tdenpos}. During the propagation of the pulse,
the width widens due to the electrons with higher energies moving faster than
the lower energy electrons, as can be seen more pronounced in the model runs
without the effect of collisions (thin curves). In the models with collisional
terms, the density decreases due to collisions of the beam particles with the
ambient plasma (thick curves). For an electron  spectrum with a higher low
energy cut off, the pulses widen due to the faster velocities of the electrons.
The beam electrons thermalised due to collisions will ultimately leave the
distribution. The total column density of electrons integrated through the simulation domain plotted
in Figure~\ref{fig:tdenvariation} shows that about 90\% of the beam electrons
are lost due to collisions within less than a second.
Although the injected electron beam pulse has a width of 160~km (0.2 millisecond after injection)
along the propagation direction, the pulse width widens as it propagates.
However the width of the radio emitting region depends on the conditions in 
the ambient medium, such as the onset of instabilities,  small-scale inhomogeneities 
in density and magnetic field [\citet{Vlasov2002}; \citet{Rozhansky2008}; \citet{Li2011a}; \citet{Li2011b}; \citet{Li2011c}; \citet{Li2012}].
In general, the temporal and spatial scales of radio emission are expected to be smaller than
that seen in the simulations reported here.  In order to compare the results of the simulations 
with the durations of the observed spikes, realistic simulations that follow the evolution of 
the waves are necessary.

\begin{figure}[h]
\hbox{
\includegraphics[width=6.5cm]{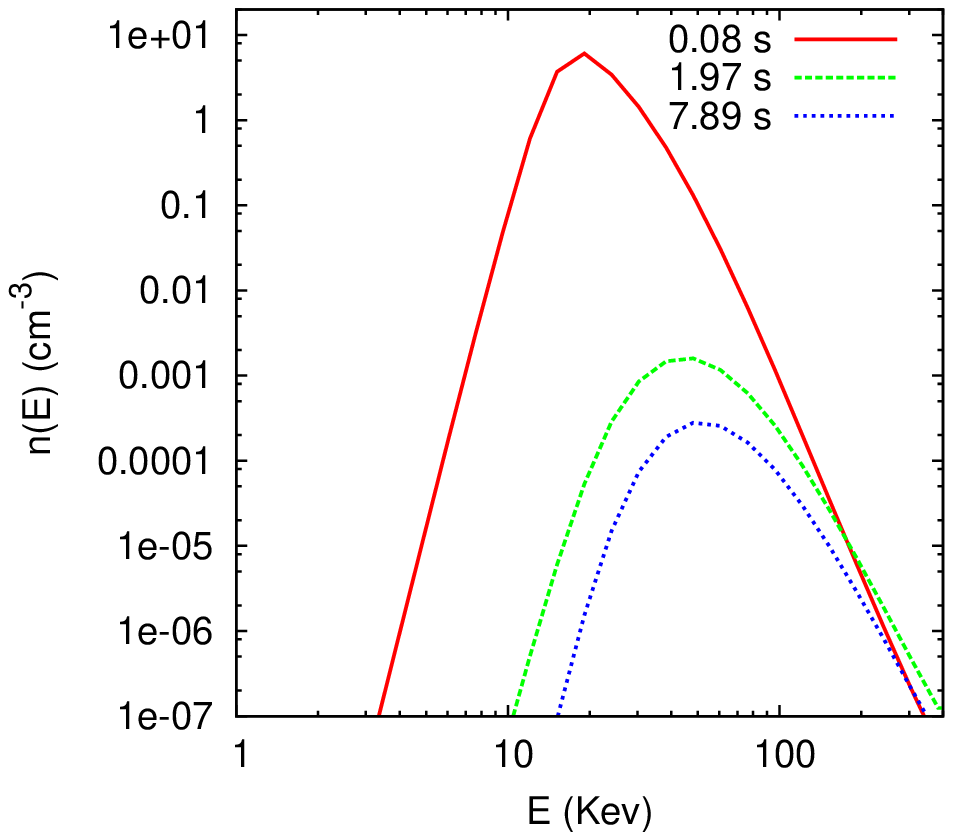}
\includegraphics[width=6.5cm]{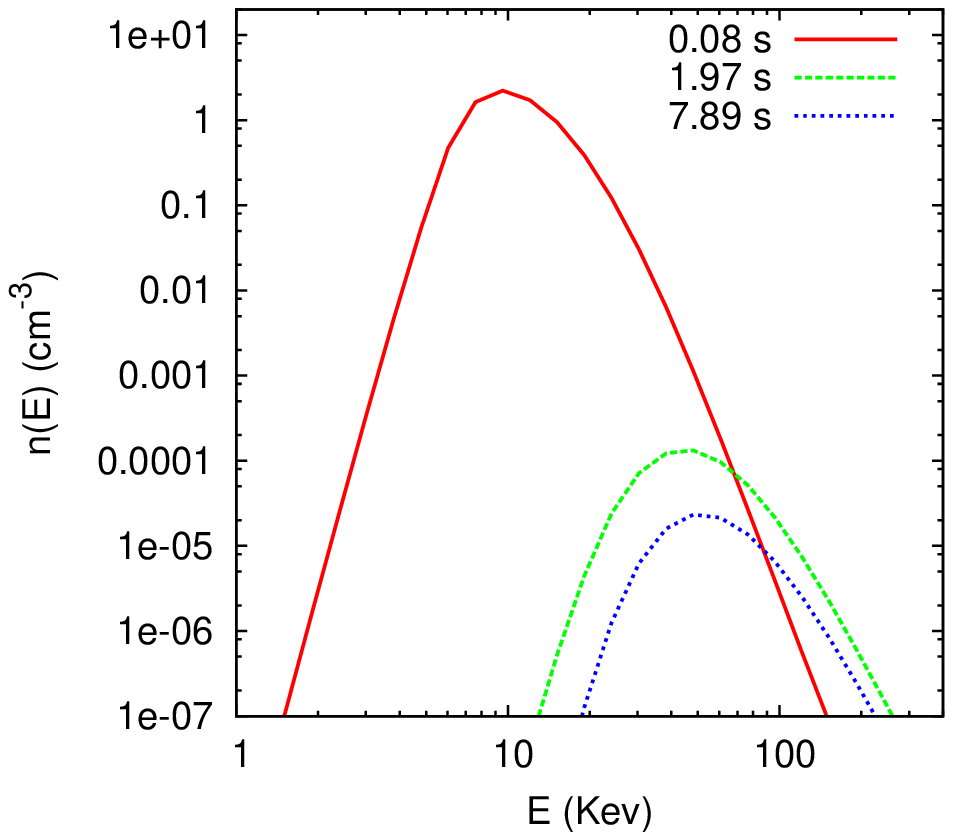}
}
\centering
\caption{Spectrum at the location where the peak of the pulse occurs is plotted for
the model runs B1 (left panel) and B2 (right panel). The solid (red), dashed (green) and 
dotted (blue) curves represent model runs at times of $0.085$, $1.97$ and $7.89$~s respectively.
At later times the peak of the spectrum shifts to higher energies due to loss of low energy electrons. 
Since the high energy electrons are fast moving, at a given location the high energy 
spectral index appears steeper than the initial injected spectral index.
}\label{fig:specevol}

\end{figure}

\begin{figure}
\hbox{
\includegraphics[width=6.5cm]{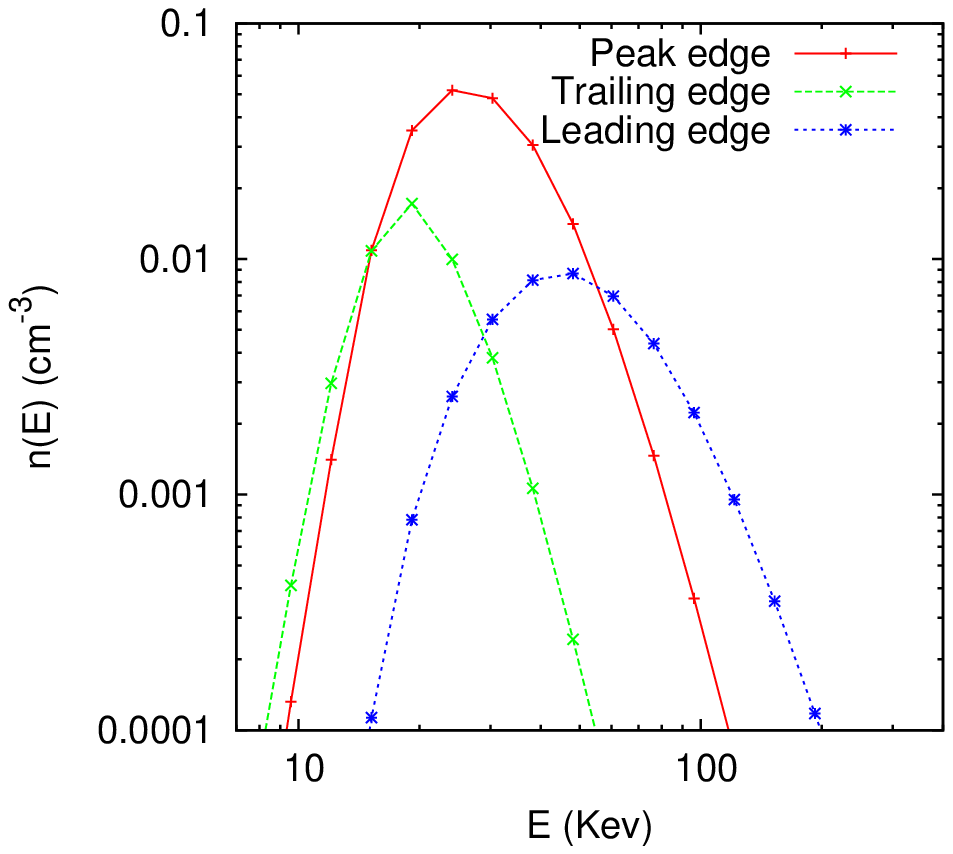}
\includegraphics[width=6.5cm]{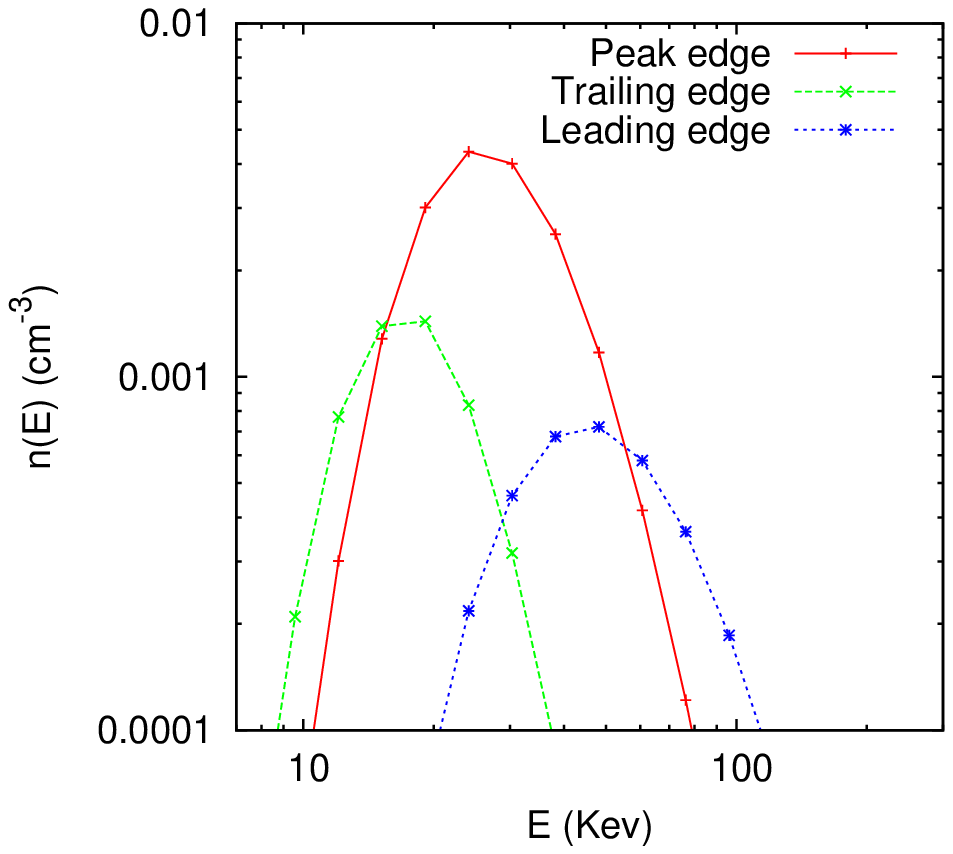}
}
\centering
\caption{Spectra at the peak (solid, red), leading (dotted, blue) and trailing (dashed, green) edges
of a pulse are plotted, for a time of $0.489$~s. 
The left and right panels correspond to model run B1 and B2 respectively.
The leading edge of the pulse mostly contains high energy electrons while the trailing 
edge contains the low energy ones.
}\label{fig:spectrapos}

\end{figure}

\subsection{Evolution of the electron spectrum}\label{s:spectrum}

In the Figure~\ref{fig:specevol}, the pitch angle averaged spectrum  
at the peak position of the pulse are plotted for different times. 
The low energy electrons are removed by collisions, thus the spectrum 
peaks at higher energies at later times. 
The fast moving electrons populate the leading edge of the pulse while the trailing edge of the 
pulse is populated by the low energy electrons as shown in Figure~\ref{fig:spectrapos}.
This can create a bump in tail distribution where $\partial{f}/\partial{v}>0$ \citep{Aschwanden2005}. 
However, in order to estimate the radio emission, evolution of the waves generated 
by the electron beam has to be calculated along with the evolution of the particle distribution.

\section{Conclusions}\label{s:conclusion}
The evolution of an electron beam pulse injected from the top of 
the coronal magnetic loop into the corona is studied. 
The energetic electrons propagate faster than the low energy 
electrons and populate the leading edge of the pulse.  
Such a configuration is expected to produce radio emission 
by several processes induced by the beam instability. 
Large fraction of the electrons in a beam of short duration 
get thermalised within a second, as compared to the 
simulation of electron beams of longer duration that propagate to 1~AU. 
In order to produce millisecond duration with narrow bandwidth radio emission, 
it is necessary to simulate the propagation of short duration pulses of 
narrow energy bands coupled to the wave dynamics.\\

%
\section*{Acknowledgements}

We thank the anonymous referees for very useful comments that have hugely improved
the presentation and have helped to correct several ambiguous statements.
We also thank the referees for bringing to our notice  earlier papers on numerical 
simulations related to this work.
Gilberto A. Casillas-P\'erez thanks IGF-UNAM and DGAPA-UNAM for the support and grant 
offered during the research project. This work was partially supported by PROMEP/103-5/07/2462
 and Conacyt CB-2009-01/130523 grants.



\end{document}